\PassOptionsToPackage{binary-units,parse-numbers=false}{siunitx}

\documentclass[conference]{IEEEtran}
\usepackage[utf8]{inputenc}

\usepackage[english]{babel}
\usepackage[english]{isodate}
\usepackage{blindtext}

\usepackage{cite}

\usepackage[acronym,smallcaps]{glossaries}
\newacronym{puf}{puf}{Physical Unclonable Function}
\newacronym{hda}{hda}{helper data algorithm}
\newacronym{nvm}{nvm}{non-volatile memory}
\newacronym{ro}{ro}{ring-oscillator}
\newacronym{sram}{sram}{static random access memory}
\glsunset{sram} 
\newacronym{fpga}{fpga}{field programmable gate array}
\glsunset{fpga} 
\newacronym{asic}{asic}{application specific integrated circuit}
\glsunset{asic} 
\newacronym{kde}{kde}{kernel density estimation}
\newacronym{pca}{pca}{principal component analysis}
\newacronym{pwc}{pwc}{pairwise comparison}
\newacronym{cots}{cots}{commercial off-the-shelf}
\newacronym{uc}{mcu}{microcontroller}
\newacronym{clb}{clb}{configurable logic block}
\glsunset{clb} 
\newacronym{snd}{snd}{standard normal distribution}
\newacronym{trng}{trng}{True Random Number Generator}
\newacronym{ctw}{ctw}{Context Tree Weighting}
\newacronym{hw}{hw}{Hamming weight}
\newacronym{hd}{hd}{Hamming distance}
\newacronym{crp}{crp}{challenge-response pair}
\newacronym{lut}{lut}{look-up table}
\glsunset{lut} 
\newacronym{ip}{ip}{intellectual property}
\glsunset{ip} 
\newacronym{ci}{ci}{confidence interval}
\newacronym{pdf}{pdf}{probability density function}
\newacronym{cdf}{cdf}{cumulative distribution function}
\newacronym{iid}{iid}{independent and identically distributed}
\newacronym{rv}{rv}{random variable}
\newacronym{clt}{clt}{central limit theorem}
\newacronym{far}{far}{false acceptance rate}
\newacronym{frr}{frr}{false rejection rate}
\newacronym{pmf}{pmf}{probability mass function}
\newacronym{ecc}{ecc}{error correcting code}
\newacronym{ind}{ind}{independent but not necessarily identically distributed}
\newacronym{nist}{nist}{national institute of standards and technology}
\glsunset{nist} 
\newacronym{mds}{mds}{maximum distance separable}

  \usepackage{graphicx}
  \usepackage{epstopdf} 
  \usepackage{color}
  \usepackage{tikz}
  \usetikzlibrary{matrix,positioning,calc,decorations.pathreplacing}

  \graphicspath{{./figures/}}
\usepackage{amsmath}
\usepackage{amssymb,amscd,upgreek,mathtools}
\usepackage[alsoload=binary,obeyall,per=fraction]{siunitx}

\newcommand{\transp}[1]{{#1}'} 

\newcommand{\secret}{\vec{s}}
\newcommand{\randnum}{\vec{r}}
\newcommand{\randspace}{\mathcal{R}}
\newcommand{\pufresp}{\vec{x}}
\newcommand{\pufspace}{\mathcal{X}}
\newcommand{\codeword}{\vec{w}}
\newcommand{\codespace}{\mathcal{W}}
\newcommand{\helperword}{\vec{y}}
\newcommand{\helperspace}{\mathcal{Y}}
\newcommand{\errorvec}{\vec{\epsilon}}
\newcommand{\errorspace}{\mathcal{E}}

\DeclareMathOperator{\prob}{P}

\makeatletter
\DeclareRobustCommand\bigop[1]{%
  \mathop{\vphantom{\sum}\mathpalette\bigop@{#1}}\slimits@
}
\newcommand{\bigop@}[2]{%
  \vcenter{%
    \sbox\z@{$#1\sum$}%
    \hbox{\resizebox{\ifx#1\displaystyle.9\fi\dimexpr\ht\z@+\dp\z@}{!}{$\m@th#2$}}%
  }%
}
\makeatother

\usepackage{cleveref}
\crefformat{equation}{(#2#1#3)}
\Crefformat{equation}{Equation~(#2#1#3)}
\crefformat{figure}{Fig.~#2#1#3}
\Crefformat{figure}{Fig.~#2#1#3}
\crefformat{table}{Tbl.~#2#1#3}
\Crefformat{table}{Tbl.~#2#1#3}
\crefformat{section}{Sec.~#2#1#3}
\Crefformat{section}{Section~#2#1#3}

\usepackage{url}

\begin{document}
%
\title{Efficient Bound for Conditional Min-Entropy\\ of Physical Unclonable Functions Beyond IID}

\bstctlcite{IEEEexample:BSTcontrol}

\author{
\IEEEauthorblockN{Florian Wilde and Christoph Frisch and Michael Pehl}
\IEEEauthorblockA{Technical University of Munich\\Munich, Germany\\\{florian.wilde, chris.frisch, m.pehl\}@tum.de}
}


%


\maketitle

\begin{figure}[b]
\vspace{-0.3cm}
\parbox{\hsize}{\em
WIFS`2019, December, 9-12, 2019, Delft, Netherlands.\\
978-1-7281-3217-4/19/\$31.00 \ \copyright 2019 IEEE.
}\end{figure}

\begin{abstract}
The remaining min-entropy of a secret generated by fuzzy extraction from a Physical Unclonable Function is typically estimated under the assumption of independent and identically distributed PUF responses, but this assumption does not hold in practice.
This work analyzes the more realistic case that the responses are independent but not necessarily identically distributed. 
For this case, we extend the (n-k) bound and a tighter bound by Delvaux et al.
In particular, we suggest a grouping bound which provides a trade off for accuracy vs computational effort.
Comparison to previous bounds shows the accuracy and efficiency of our bound.
We also adapt the key rank (a tool from side-channel analysis) to cross-validate the state-of-the-art and our proposed min-entropy bounds based on publicly available PUF data from real hardware.

\end{abstract}


%
\IEEEpeerreviewmaketitle

\section{Introduction}\label{sec:intro}
With the demand for security in low-cost devices, \glspl{puf} raised attention as a cheap and still secure alternative for key storage compared to secured \gls{nvm}. 
\glspl{puf} utilize variations of the manufacturing process.
These variations cause unpredictable and uncontrollable differences in the behavior of identical circuits on different positions on a certain chip and on different chips.
The \gls{puf} response, a chip-unique secret which is not permanently stored but derived by the \gls{puf} on demand, is obtained from such differences.
This work focuses on \glspl{puf} comprised of multiple \gls{puf} cells, each contributing one bit to the \gls{puf} response.
The \emph{expected} value of a chip's \gls{puf} response is fixed during the manufacturing process.
Its value, however, may vary between measurements due to noise, environmental effects, and aging.
Hence the \gls{puf} response cannot be used directly as a key.
The typical approach to obtain a key from a \gls{puf} is to
(i) measure the noisy \gls{puf} response,
(ii) map it to a noisy codeword of an \gls{ecc} by means of helper data and a \gls{hda}, and
(iii) decode via \gls{ecc} to derive a stable key.

The helper data may be stored unprotected and are generated during a roll-out phase, usually in a code-offset fuzzy extractor~\cite{Dodis2004} or fuzzy commitment~\cite{Juels1999} scheme.
This schemes generate helper data~\mbox{$\helperword=\pufresp\oplus\codeword$} by encoding a random \SI{k}{bit}~vector~$\randnum$ to an \SI{n}{bit}~codeword~$\codeword$ and XORing an \SI{n}{bit}~\gls{puf}~response~$\pufresp$ to $\codeword$.
Depending on the scheme, the secret $\secret$, which is hashed to the actual key, is either $\pufresp$ or $\randnum$.
For performance reasons the key is usually split into multiple chunks, so an \gls{ecc} with smaller message length can be used.

For the schemes and under the assumption of bias-free and \gls{iid} \gls{puf} responses, the entropy of the key, when neglecting the hash, is limited by $k$ \cite{Pehl2017}. 
Most real world \glspl{puf} do not produce bias-free responses, though; even worse: they do not even fit to an \gls{iid} assumption.
Despite great efforts to design \gls{puf} cells for unbiased output, incautious layout, unexpected influence of adjacent logic, or small deviations in mask production can already lead to biased responses.
Because these effects differ for each position on the die, the resulting bias is not identical.
\begin{figure}
\centering\input{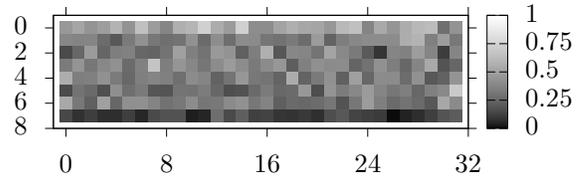}
\caption{Relative frequency of 1 (Bit-Alias) for the \gls{puf} bits derived from the dataset \cite{MaitiDataset}. Squares reflect the positions of \gls{puf} cells (\gls{ro} pairs) on a device.}\label{fig:MaitiHeatmap}
\end{figure}
An example is shown in \cref{fig:MaitiHeatmap}.
The heat map shows for each cell of an \gls{ro} \gls{puf} the relative frequency $p$ for the corresponding response bit to turn out as a 1 among all devices in the dataset (i.e. the Bit-Alias \cite{Maiti2010}).
The underlying \gls{ro} frequencies are taken from \cite{MaitiDataset}; a response bit is derived by mutually exclusive pairwise comparions between two adjacent \glspl{ro} in a row.
Previous work shows that this approach produces responses almost entirely free of spatial correlation \cite{Wilde2018}, but cannot remove the observable bias, which can be explained by placement and routing \cite{Pehl2016}.
Similar effects with varying impact are frequently observable in \gls{puf} implementations.

An attacker who tries to reveal the key stored by a \gls{puf} knows the helper data and the type of \gls{ecc}.
Thus, leveraging the Bit-Aliases gives the attacker a significant advantage when guessing the key.
The effort for guessing the key in such a setting might be approximated by guesswork and its relatives \cite{Massey1994, Pliam2000, NISTsp800-90B}.
But this is not feasible for the \gls{puf} setting due to its high computational complexity.
Another tool to bound the quality of a key is entropy. 
Due to the cryptographic relevance, only min-entropy is considered throughout this work.

\emph{Contribution:}
Estimation of the key entropy given known helper data was previously studied in \cite{Dodis2004,Delvaux2015efficient}.
However, these works only cover the case of biased \gls{iid} and unbiased correlated \gls{puf} responses. 
This work first summarizes the main achievements in the state of the art and demonstrates the problems, which are caused when (i) not considering the blockwise processing during post-processing, or (ii) assuming \gls{iid} instead of \gls{ind} \gls{puf} responses.
Several improvements upon previous approaches are presented.
In particular, we introduce the \emph{grouping bound}, which provides a practically feasible bound of the conditional min-entropy of \gls{ind} \gls{puf} responses. 
All improvements are demonstrated using real world \gls{puf} data.
Additionally, the entropy estimates are compared to the key rank, i.e. the actual effort for guessing a key.

\emph{Structure:}
After discussing the state of the art of entropy estimation for \glspl{puf} in \cref{sec:sota}, we introduce our improvements in \cref{sec:entropyIND}.
Results are provided and analyzed in the context of key rank in \cref{sec:keyrank}. 
\cref{sec:conclusion} concludes our work.

\section{State-of-the-art Entropy Estimation for \glspl{puf}}\label{sec:sota}
To evaluate the security of a \gls{puf} based key storage, 
the \emph{conditional entropy} of the \gls{puf} response $\pufresp$ given the helper data $\helperword$ is the important figure of merit.
This section presents state-of-the-art methods to calculate or bound the conditional entropy, but first discusses entropy estimation of the \gls{puf} alone as a necessary prerequisite.

\subsection{\gls{puf} without Helper Data}
Typically, response bits of \glspl{puf} are considered realizations of \gls{iid} Bernoulli random variables.
I.e. each bit takes the value~1 with probability $p$ independent of all other bits.
Under this assumption, and given a sufficiently large number of observed bits, if necessary from multiple devices, $p$ is well approximated by the relative frequency of a 1 in the data set.
Then the \emph{min-entropy} $m$ of a device's \SI{n}{\bit} \gls{puf} response $\pufresp$ is
\begin{equation}\label{eq:pufiid}
  m = H_\infty(X) = -n \log_2 \left( \max\left(p, 1-p\right)\right) \text{.}
\end{equation}
However, as already pointed out in \cref{sec:intro}, for many \gls{puf} implementations the probability for a 1 differs significantly between the positions on the die as shown in \cref{fig:MaitiHeatmap}.
This contradicts the \gls{iid} assumption so that \cref{eq:pufiid} becomes imprecise.

\begin{figure}
\centering\input{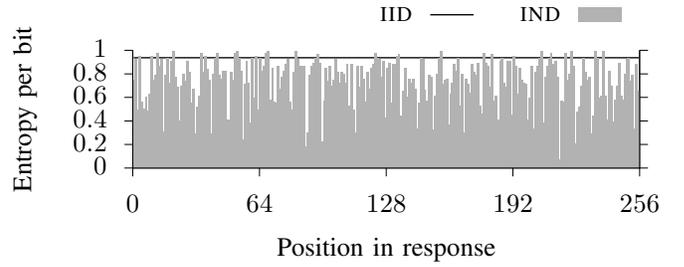}
\caption{Min-entropy per bit for \acrshort{puf} response $\pufresp$ under \acrshort{iid} assumption and for \acrshort{ind} case.
\acrshort{iid} uses \cref{eq:pufiid} without multiplication by $n$ and with $p=\sum p_i/n$.
\acrshort{ind} uses \cref{eq:pufniid} without summation.}\label{fig:PUFbitEntropy}
\end{figure}

Consequently, Wilde et al. \cite{Wilde2014} assumed the bits of the \gls{puf} responses to be \gls{ind}.
In this case, the probabilities for a 1 form a vector\footnote{All vectors in this work are row vectors unless explicitly stated otherwise.} $\vec{p}=\begin{pmatrix}p_1 & \cdots & p_n\end{pmatrix}$, where $p_i$ is approximated by the relative frequency of a 1 among the bits originating from position $i$ on the die.
The min-entropy under the \gls{ind} assumption is
\begin{equation}\label{eq:pufniid}
  \tilde{m} = \tilde{H}_\infty(X) = -\sum_{i=1}^{n} \log_2 \left( \max\left(p_i, 1-p_i\right)\right) \text{.}
\end{equation}
The extent of the variation in entropy per bit, together with the approximated entropy per bit under the \gls{iid} assumption, is depicted in \cref{fig:PUFbitEntropy}.
For most positions, the entropy under \gls{iid} assumption is severely overestimated compared to the \gls{ind} case, which sums up to \SI{240}{\bit} instead of \SI{188}{\bit} over all positions, cf. \cref{tbl:entropies}.
The root cause for the difference in entropy estimates is visible in \cref{fig:MaitiHeatmap}:
Only few positions have $p_i \approx 0.5$, while most positions suffer from either $p_i>0.5$ or $p_i<0.5$, i.e. a bias towards 1 or towards 0.
Both cases cause a reduction in entropy $\tilde{m}$ according to \cref{eq:pufniid}. 
In the \gls{iid} case, however, $p$ equals the average over all $p_i$, which causes those $p_i>0.5$ to partially cancel out with those $p_i<0.5$, resulting in $p$ being incorrectly close to $0.5$ and overestimating entropy.

\subsection{(n-k) Bound}
Given the entropy of the \gls{puf} alone, an easy-to-compute and frequently used, yet pessimistic, lower bound for the conditional entropy of the \gls{puf} response given the helper data is the (n-k) bound \cite{Dodis2004}.
It applies only to the \gls{iid} case, where it calculates the entropy estimate $l$ of an \SI{o}{\bit} key $\secret$ by a simple balance sheet:
Ingoing to the fuzzy extractor are \SI{m}{\bit} of entropy from the \gls{puf} response $\pufresp$ and \SI{k}{\bit} of entropy from a random number $\randnum$ that has full entropy.
Outgoing are \SI{n}{\bit} of entropy for the helper data $\helperword$.
An additional loss of \SI{L}{\bit} occurs in the hash function that compresses the fuzzy extractor output into a fixed length key.
\begin{equation}\label{eq:nkbound}
    l=m+k-n-L
\end{equation}
Neglecting the hash function, the entropy loss between $\pufresp$ and $\secret$ is $n-k$, which gave the bound its name.

This approach can result in negative values for $l$, cf. \cref{tbl:entropies}, because $n-k$ is only an upper bound for the entropy loss that holds with equality only if $m=n$ \cite{Delvaux2015efficient}.

\subsection{Average Conditional Min-Entropy}
Instead of the pessimistic (n-k) bound, Delvaux et al. \cite{Delvaux2015efficient} provide a direct mathematical expression for the conditional entropy based on the code-offset fuzzy extractor described by Dodis et al. \cite{dodis2008fuzzy}.
Evaluation is only feasible for small codes \cite{Delvaux2015efficient}, but it provides a baseline to compare the bounds against.
By application of Bayes' rule to the \emph{average conditional min-entropy} definition in \cite{dodis2008fuzzy} they yield
\begin{equation}
  H_\infty \left(X|Y\right) = -\log_2 \left(\frac{1}{|\randspace|} \sum_{\helperword \in \helperspace} \max\limits_{\codeword\in\codespace} \prob\left(X=\helperword \oplus \codeword\right)\right) \label{eq:Delvaux8}
\end{equation}
for an \gls{ecc} in general and respectively the more efficient
\begin{equation}
  H_\infty \left(X|Y\right) = -\log_2 \left(\sum_{\errorvec\in\errorspace} \max\limits_{\codeword\in\codespace} \prob\left(X=\errorvec \oplus \codeword\right) \right)  \label{eq:Delvaux9}
\end{equation}
for linear \glspl{ecc}.
Here $\randspace = \{0,1\}^{1\times k_b}$ is the set of possible messages of the \gls{ecc} and $\helperspace = \{0,1\}^{1\times n_b}$ the set of possible helper data generated for one \gls{ecc} block.
$\codeword$ is a codeword from the codeword space $\codespace$ and $\errorvec$ an element of the minimum Hamming-weight error vector space $\errorspace$ such that $\{\codeword\oplus\errorvec\ | \codeword\in\codespace, \errorvec\in\errorspace\} = \helperspace$, i.e. $\errorvec$ is a coset leader.
Using a 3-repetition code as example, $\mathcal{W} = \{000, 111\}$ and $\mathcal{E} = \{000, 001, 010, 100\}$.

\Cref{eq:Delvaux8,eq:Delvaux9} are not limited to block codes \cite{Delvaux2015efficient}, but in such a case, they are applied on a single \gls{ecc} block only.
For the \gls{iid} case considered in \cite{Delvaux2015efficient}, it is sufficient to multiply the result with the number of \gls{ecc} blocks required to produce the entire secret $\secret$ to get an overall result.

\subsection{Efficient Lower Bound for Average Conditional Min-Entropy for Large $n_b$}
Because \cref{eq:Delvaux8}, \cref{eq:Delvaux9} require up to $2^{n_b}|\randspace|$ or respectively $2^{n_b}$ operations, direct evaluation even for moderately sized codes such as a (63,7,15) \textsc{bch}-code is hardly feasible.
Therefore, \cite{Delvaux2015efficient} proposed an algorithm that reduces computational effort via groups of \gls{puf} responses with the same probability, sorted in decreasing order of probability, to avoid computations that will later be discarded by the $\max$ operator in \cref{eq:Delvaux8}, \cref{eq:Delvaux9}.

Under the \gls{iid} assumption, all $2^{n_b}$ responses can be grouped into $J=n_b+1$ groups $\varphi_j$ with probability
\begin{equation}
  q_j = q^j (1-q)^{n_b-j}
\end{equation}
where $j \in [0, n_b]$ and $q = \min(p, 1-p)$.
Each $\varphi_j$ then contains
\begin{equation}
  |\varphi_j| = \binom{n_b}{j}\label{eq:cardPhiIID}
\end{equation}
\gls{puf} responses $\pufresp$ with probability $q_j$ and
\begin{equation}
  q_j > q_{j+1}\text{.}\label{eq:phiDescendProb}
\end{equation}

\begin{figure}
  \centering
  \tikzset{helpermatrix/.style={matrix of nodes,row sep=1ex,column sep=1ex,nodes={inner sep=0.5ex}}}
  \tikzset{pufmatrix/.style={helpermatrix}}
  \tikzset{randmatrix/.style={helpermatrix,nodes={text width=3.3ex,align=center}}}
  \begin{tikzpicture}[node distance=1ex]
    \matrix (Y1) [helpermatrix] {
      000 & 001 & 010 & 100\\
      111 & 110 & 101 & \textbf{011}\\
    };
    \matrix (Y2) [helpermatrix,right=2ex of Y1] {
      001 & 010 & 100 & 000\\
      110 & 101 & 011 & 111\\
    };
    \node [left=of Y1] {$\helperword$};
    \draw [dashed] (Y1.north west) rectangle (Y1.south east) node [xshift=-0.6ex,fill=white,inner sep=2pt] {$\helperspace$};
    \draw [dashed] (Y2.north west) rectangle (Y2.south east) node [xshift=-0.6ex,fill=white,inner sep=2pt] {$\helperspace$};
    \matrix (X1) [pufmatrix,above=of Y1] {
      000 & 001 & 010 & \textbf{100}\\
    };
    \matrix (X2) [pufmatrix,above=of Y2] {
      110 & 101 & 011 & 111\\
    };
    \node [left=of X1] {$\pufresp$};
    \draw [dashed] (X1.north west) rectangle (X2.south east) node [yshift=1pt,fill=white,inner sep=2pt] {$\pufspace$};
    \draw [decorate,decoration={brace,amplitude=1ex}] ($ (X1-1-1.north west) + (0ex,2ex) $) -- ($ (X1-1-1.north east) + (0ex,2ex) $) node [midway,above,yshift=1ex] {$\varphi_0$};
    \draw [decorate,decoration={brace,amplitude=1ex}] ($ (X1-1-2.north west) + (0ex,2ex) $) -- ($ (X1-1-4.north east) + (0ex,2ex) $) node [midway,above,yshift=1ex] {$\varphi_1$};
    \draw [decorate,decoration={brace,amplitude=1ex}] ($ (X2-1-1.north west) + (0ex,2ex) $) -- ($ (X2-1-3.north east) + (0ex,2ex) $) node [midway,above,yshift=1ex] {$\varphi_2$};
    \draw [decorate,decoration={brace,amplitude=1ex}] ($ (X2-1-4.north west) + (0ex,2ex) $) -- ($ (X2-1-4.north east) + (0ex,2ex) $) node [midway,above,yshift=1ex] {$\varphi_3$};
    \matrix (C1A) [pufmatrix,below=2ex of Y1-2-1,nodes={fill=black!10}] {
      000\\
      111\\
    };
    \matrix (C1B) [pufmatrix,below=2ex of Y1-2-4,nodes={fill=black!10}] {
      000\\
      \textbf{111}\\
    };
    \node at ($ 0.5*(C1A) + 0.5*(C1B) $) {$\cdots$};
    \node [left=of C1A] {$\codeword$};
    \draw [dashed] (C1A.north west) rectangle (C1A.south east) node [xshift=0.9ex,yshift=0.5ex,fill=white,inner sep=1pt] {$\codespace$};
    \draw [dashed] (C1B.north east) rectangle (C1B.south west) node [xshift=-0.9ex,yshift=0.5ex,fill=white,inner sep=1pt] {$\codespace$};
    \matrix (C2A) [pufmatrix,below=2ex of Y2-2-1] {
      111\\
      000\\
    };
    \matrix (C2B) [pufmatrix,below=2ex of Y2-2-4] {
      111\\
      000\\
    };
    \node at ($ 0.5*(C2A) + 0.5*(C2B) $) {$\cdots$};
    \draw [dashed] (C2A.north west) rectangle (C2A.south east) node [xshift=0.9ex,yshift=0.5ex,fill=white,inner sep=1pt] {$\codespace$};
    \draw [dashed] (C2B.north east) rectangle (C2B.south west) node [xshift=-0.9ex,yshift=0.5ex,fill=white,inner sep=1pt] {$\codespace$};
    \draw [decorate,decoration={brace,amplitude=1ex,mirror}] ($ (C1A.south west) - (0ex,1ex) $) -- ($ (C1B.south east) - (0ex,1ex) $) node [midway,below,yshift=-1ex] {1\textsuperscript{st} guess};
    \draw [decorate,decoration={brace,amplitude=1ex,mirror}] ($ (C2A.south west) - (0ex,1ex) $) -- ($ (C2B.south east) - (0ex,1ex) $) node [midway,below,yshift=-1ex] {2\textsuperscript{nd} guess};
    \draw [dotted] ($ 0.5*(X1.north east) + 0.5*(X2.north west) + (0ex,4ex) $) -- ($ 0.5*(C1B.south east) + 0.5*(C2A.south west) - (0ex,5ex) $);
    \draw (Y2-1-4.east) -- ++(3ex,0ex) node (rel1) [fill=white,inner sep=0pt] {$\oplus$};
    \draw (C2B-1-1.east) -| (rel1) |- (X2-1-4.east) node [midway,coordinate] (rel1e) {};
    \draw (Y2-2-4.east) -- ++(5ex,0ex) node (rel2) [fill=white,inner sep=0pt] {$\oplus$};
    \draw (C2B-2-1.east) -| (rel2) |- (rel1e);
  \end{tikzpicture}
  \caption{Visualization of the algorithm from \cite{Delvaux2015efficient} using a 3-repetition code and $p<0.5$.
  The best guess -- which is the one selected by the $\max$ operator in \cref{eq:Delvaux8}, \cref{eq:Delvaux9} -- for every possible $\helperword$ can be found in the left half of the figure.
  E.g. for $\helperword=011$, the best guess is to assume $\pufresp=100$ thus $\codeword=111$, $\randnum=1$.}\label{fig:delvauxBlocks}
\end{figure}
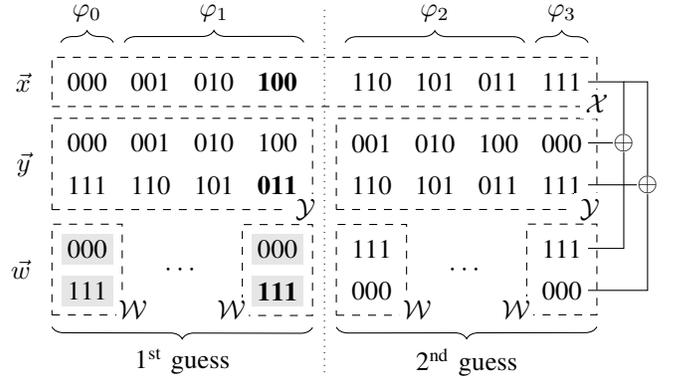

This is useful because, for a given $\helperword$, the $\max$ operator in \cref{eq:Delvaux8}, \cref{eq:Delvaux9} selects the most likely $\pufresp$ that is reachable by addition of a $\codeword$.
Therefore, and because $\randnum$ is chosen uniformly, $|\randspace|=|\codespace|$ elements of $\helperspace$ map to the same $\pufresp$ as best guess, cf.~\cref{fig:delvauxBlocks}.
E.g. for $\helperword=100$ and $\helperword=011$, $\pufresp=100$ is the most likely $\pufresp$ within reach via $\codeword=000$, $\codeword=111$.
It is thus sufficient to consider $2^{n_b}/|\randspace|$ (or $2^{n_b-k_b}$ for block codes) elements of $\pufspace$ to accumulate the best guesses for all $\helperword \in \helperspace$.

\emph{Which} $2^{n_b}/|\randspace|$ elements of $\pufspace$ to consider depends on the specific \gls{ecc}.
For repetition codes with odd $n_b$, $\bigcup_{j=0}^t \varphi_j$, i.e. the $2^{n_b}/|\randspace|$ \emph{most likely} $\pufresp$, are the correct choice, because due to the \gls{iid} assumption they either equal $\errorspace$ for $p<0.5$, cf. \cref{fig:delvauxBlocks}, or, if $p>0.5$, can be mapped to $\errorspace$ by addition of a fixed $\codeword$.
However, for a (15,5,3) \textsc{bch}-code, in addtion to $\bigcup_{j=0}^3 \varphi_j$, 420 $\pufresp$ from $\varphi_4$ and 28 from $\varphi_5$ are required.
By instead choosing 448 $\pufresp$ from $\varphi_4$, thus again the $2^{n_b}/|\randspace|$ \emph{most likely} $\pufresp$, we overestimate the probability for 28 out of 1024 $\pufresp$.
\emph{Always} choosing the $2^{n_b}/|\randspace|$ most likely $\pufresp$ thus provides an upper bound for the sum in \cref{eq:Delvaux8}, \cref{eq:Delvaux9}, which leads to a lower bound for $H_\infty \left(X|Y\right)$ that holds with equality for \gls{mds} \glspl{ecc}.

As $q_j$ is equal for all $\pufresp$ in a $\varphi_j$, $|\varphi_j|$ responses $\pufresp$ can be processed at once.
Because $|\varphi_j|$ often exceeds $|\randspace|$ already for small $j$, this provides the most speed up, so that for a linear ($n_b$,$k_b$,$t$) block code, $t$ to $t+1$ computations suffice.

To summarize, this approach results in a much tighter lower bound for the remaining min-entropy than the (n-k) bound and is feasible in practice if the \gls{iid} assumption holds.

\begin{table*}
\caption{estimations and bounds for min-entropy of $\vec{x}$ and for conditional min-entropy of $\vec{s}$ for (n)-rep and (n,k,t)-bch codes}\label{tbl:entropies}
\centering
\begin{tabular}{l
                S[table-format=3]
                S[table-format=3]
                S[table-format=3]
                S[table-format=3]
                S[table-format=3.2]
                S[table-format=-2.1]
                S[table-format=2.2]
                S[table-format=3.1]
                S[table-format=2.1]
                S[table-format=2.2]
                S[table-format=2.2]}
                &   &   &   &   & \multicolumn{3}{|c}{(n-k) bound} & \multicolumn{2}{|c}{avg. cond. min-entropy} & \multicolumn{2}{|c}{Grouping, $\theta_\Delta=$} \\
 & \multicolumn{1}{c}{$n$} & \multicolumn{1}{c}{$m$} & \multicolumn{1}{c}{$\tilde{m}$} & \multicolumn{1}{c}{$k$} & \multicolumn{1}{c}{$l$} & \multicolumn{1}{c}{$l (\tilde{m})$} & \multicolumn{1}{c}{$\tilde{l}$} & \multicolumn{1}{c}{\acrshort{iid}} & \multicolumn{1}{c}{\acrshort{ind}} & 0.05 & 0.10 \\
\hline
 & 256 & 240 & 188 &  &  &  &  &  & \\
(3) & 255 & 239 & 187 &  85 & 69.0 & 17.3 & 22.4 & 77.1 & 47.2 & 45.9 & 43.9 \\
(5) & 255 & 239 & 187 &  51 & 35.0 & -16.7 & 3.30 & 45.1 & 22.5 & 21.7 & 19.6 \\
(7) & 252 & 236 & 185 &  36 & 20.2 & -30.7 & 0.01 & 31.2 & 12.5 & 11.9 & 10.6 \\
(21) & 252 & 236 & 185 &  12 & -3.78 & -54.7 & 0.00 & 9.38 & 1.56 & 1.25 & 0.86 \\
(7,4,1) & 252 & 236 & 185 & 144 & 128 & 77.3 & 77.3 & 132 & 88.4 & 85.1 & 79.5 \\
(15,5,3) & 255 & 239 & 187 &  85 & 69.0 & 17.3 & 17.3 & 76.2 & 40.6 & 37.3 & 33.0 \\
(31,6,7) & 248 & 232 & 182 &  48 & 32.5 & -18.1 & 0.00 & NaN & NaN & 13.3 & 10.7 \\
(63,7,15) & 252 & 236 & 185 &  28 & 12.2 & -38.7 & 0.00 & NaN & NaN & 3.82 & 2.36 \\
(127,8,31) & 254 & 238 & 186 &  16 & 0.09 & -51.6 & 0.00 & NaN & NaN & 0.57 & 0.19 \\

\end{tabular}
\end{table*}

\section{Improved Blockwise Conditional Entropy Estimations for the \acrshort{ind} Case}\label{sec:entropyIND}
In this section, we present improved entropy bounds for keys derived from \glspl{puf}.
They are developed under the constraint of practical applicability and consider that \gls{puf} response bits are (i) \gls{ind} rather than \gls{iid} and (ii) usually processed blockwise.

\subsection{(n-k) Bound}
To bring the (n-k) bound to the \gls{ind} case, $m$ is to be replaced by $\tilde{m}$ from \cref{eq:pufniid}.
However, this only uses a better estimation for the ingoing entropy from the \gls{puf}, while the entropy loss is still at its worst-case.
Hence the exact entropy according to \cref{eq:Delvaux8} is significantly underestimated, cf. \cref{tbl:entropies}, which might cause overdesign and, therefore, high costs.

A tighter, yet very fast to calculate bound is found when considering blockwise processing of \gls{puf} bits. 
The blocks are independent of each other even in the \gls{ind} case.
Therefore, one block cannot leak information about another and entropy can be estimated per block.
\cref{fig:entropyNK} shows the estimated entropy per bit given $p_i$ values according to \cref{fig:MaitiHeatmap};
bounds are calculated for the complete \gls{puf} response under the \gls{iid} and the \gls{ind} assumption as well as blockwise assuming \gls{ind} and a 5-repetition code.
The actual entropy per block is obviously bound from below by zero.
Thus, the blockwise estimate of the overall entropy is
\begin{equation}\label{eq:nkind}
  \tilde{l} = \sum_{i=1}^{N_b } \max(\tilde{m}_i + k_b - n_b, 0)
\end{equation}
where $\tilde{m}_i$ is the entropy in the \gls{puf} bits processed in block~$i$;
the number of blocks for encoding a complete key is \mbox{$N_b=\lfloor \frac{n}{n_b}\rfloor$}.
Each block encodes $k_b$ disjoint bits from the random number $\randnum$, which is assumed to have $k=k_b\cdot N_b$ bits of entropy, and contributes $n_b$ bits of helper data to $\helperword$.
\cref{tbl:entropies} shows the improvement through the tightened bound for real world data \cite{MaitiDataset} and different codes.
\begin{figure}
\centering\input{figures/entropyNK.tex}
\caption{Min-entropy per block for $\secret$ for a 5-repetition code using original (n-k) bound and improved version \cref{eq:nkind}.
For \acrshort{iid}, i.e. using $m$ in \cref{eq:nkbound}, $l/N_b$ is overestimated; $l(\tilde{m})/N_b$ is a too pessimistic bound.
\acrshort{ind} shows the entropy estimate per block $\tilde{m}_i+k_b-n_b$.}\label{fig:entropyNK}
\end{figure}

\subsection{Average Conditional Min-Entropy}
Under the \gls{iid} assumption, it is sufficient to apply \cref{eq:Delvaux8} or \cref{eq:Delvaux9} on one block.
The exact entropy in the key is then derived by multiplying the result with the number of required codewords~$N_b$. 
To extend this exact approach to the \gls{ind} case, we  compute the entropy individually for all $N_b$ blocks, using the accurately estimated probabilities in $\vec{p}$ to get $\prob\left(X=\errorvec \oplus \codeword\right)$.
The entropies per block derived  with \cref{eq:Delvaux9} under the \gls{iid} and \gls{ind} assumption are depicted in \cref{fig:entropyDVX} and listed in \cref{tbl:entropies}. 
Both show that falsely making an \gls{iid} assumption severely overestimates the remaining entropy, which puts the security of the overall system at risk.
\begin{figure}
\centering\input{figures/entropyDVX.tex}
\caption{Min-entropy per block for $\secret$ for a 5-repetition code using original and blockwise average conditional min-entropy.
As with the (n-k) bound, the \acrshort{iid} assumption severely overestimates entropy compared to the \acrshort{ind} case.}\label{fig:entropyDVX}
\end{figure}

\subsection{The Grouping Bound: An Efficient Lower Bound for Average Conditional Min-Entropy for Large $n_b$}
For the \gls{ind} case, which is not considered in \cite{Delvaux2015efficient}, no two responses might have equal probability.
If so, grouping according to \cite{Delvaux2015efficient} ends up in $2^{n_b}$ groups $\varphi_j$ with $|\varphi_j|=1$.
Consequently, the final reduction from $2^{n_b}/|\randspace|$ to $t$ or $t+1$ computations is no longer applicable, which makes computation infeasible for large codes.
In the following, we propose a new algorithm to calculate a strict lower bound (or optionally an approximation) for the average conditional min-entropy for the \gls{ind} case in feasible time.

First, assume that for all $p_i$ in $\vec{p}$ 
\begin{equation}
    \forall i : p_i \geq 0.5. \label{eq:all1bias}
\end{equation}
Practically, this is achieved by setting each $p_i$ that violates the assumption to $1-p_i$ and inverting the corresponding helper data bit.
Because $H_\infty(p) = H_\infty(1-p)$, entropy is unaffected and the entropy estimation obtained from the preprocessed data holds for the original data.
For the special case where, after the transform, $\exists 0\leq p_a \leq 1 : (\forall i: (p_i \approx p_a))$, which fits for some \gls{sram} \glspl{puf} \cite{Wilde2017xmc}, sufficiently accurate results may be obtained under the \gls{iid} assumption with $p$ set to $p_a$.

For the general \gls{ind} case, though, a method is needed to identify and calculate the probability of the $2^{n_b}/|\randspace|$ most likely \gls{puf} responses $\pufresp$ \emph{efficiently}.
The key to this are large response groups $\varphi_j$, because all $\pufresp$ in a group $\varphi_j$ can be covered by a single computation.
$|\varphi_j|$ is strongly affected by the number of unique values in $\vec{p}$, because if some $p_i$ are equal, the probability of an $\pufresp$ only depends on how many, but not which, of the corresponding response bits are flipped in $\pufresp$, creating a large number of $\pufresp$ to be put into the same $\varphi_j$.
Due to the fast increase of the binomial coefficient, this gain is already large for a small number of equal $p_i$.

We therefore suggest to trade off accuracy against computational effort:
The $p_i$ are collected in $T$ bias groups $\rho_\tau$ such that the difference between any two $p_i$ within any $\rho_\tau$ is at most $\theta_\Delta$.
All $\eta_\tau$ $p_i$ within a group $\rho_\tau$ are assigned a representative probability $\theta_\tau$ and $\vec{\theta} = \begin{pmatrix} \theta_1 & \cdots & \theta_T\end{pmatrix}$.
The value of $\theta_\tau$ may be the maximum of all $p_i$ in the group to obtain a strict lower bound for the remaining entropy, or some kind of average, e.g. mean or median, to obtain an approximation of the remaining entropy.
Changing $\theta_\Delta$ trades off tightness of the bound against computational cost, because it equals the maximum error made by approximation of some $p_i$ and influences $|\varphi_j|$ via $\eta_\tau$.
Independent of the chosen $\theta_\Delta$, all response bits have to belong to exactly one bias group $\rho_\tau$, thus
\begin{equation}
  n_b = \sum_{\tau=1}^T \eta_\tau\text{.}
\end{equation}

After the transition from $\vec{p}$ to $\vec{\theta}$, the probability of an $\pufresp$ only depends on the number of flipped bits from each bias group $\rho_\tau$.
Hence, a response group $\varphi_j$, which contains all $\pufresp$ with same probability, is uniquely described by the number $\zeta_{j,\tau}$ of bits flipped from each group $\rho_\tau$ or by a flip vector $\vec{\zeta_{j}}=\begin{pmatrix}\zeta_{j,1} & \cdots & \zeta_{j,T}\end{pmatrix}$.
Assume this vectors to be stored in a $J \times T$ matrix
\begin{equation}
  \check{Z}=\transp{\begin{pmatrix}\transp{\vec{\zeta}_{0}} & \cdots & \transp{\vec{\zeta}_{J}}\end{pmatrix}}\text{,}
\end{equation}
which is sorted according to \cref{eq:phiDescendProb}.
Transposition of a vector is indicated in this work by a$\transp{~}$.
The number of rows in $\check{Z}$ then equals the number of response groups $\varphi_j$,
\begin{equation}
  J = \prod_{\tau=1}^T (\eta_\tau+1)\text{,}
\end{equation}
and each group $\varphi_j$ contains
\begin{equation}
  |\varphi_j| = \prod_{\tau=1}^T \binom{\eta_\tau}{\zeta_{j,\tau}}\label{eq:groupingCardinality}
\end{equation}
\gls{puf} responses $\pufresp$ with same probability $q_j$.
We summarize this group sizes as $\vec{\psi} = \begin{pmatrix}|\varphi_0| & \cdots & |\varphi_J|\end{pmatrix}$.
Because of \cref{eq:all1bias}, the best guess is $\pufresp=\vec{1}=\begin{pmatrix} 1 & \cdots & 1\end{pmatrix}$, which corresponds to $\vec{\zeta_0}=\vec{0}$, and
\begin{equation}
  q_0 = \prod_{\tau=1}^T \theta_\tau^{\eta_\tau}\text{,}
\end{equation}
which can be written in the log-domain as
\begin{equation}
  \varsigma_0 = \log_2\left(q_0\right) = \vec{\eta} \log_2\transp{\left(\vec{\theta}\,\right)}
\end{equation}
using a scalar product of two vectors.
Based on $\varsigma_0$, the $\log$-probabilities $\varsigma_j$ of all response groups $\varphi_j$ follow from 
\begin{equation}
  \transp{\left.\vec{\varsigma}\,\right.} = \transp{\left.\vec{1}\right.} \varsigma_0 + \check{Z} \transp{\left(\log_2\left(\vec{1} - \vec{\theta}\,\right) - \log_2\left(\vec{\theta}\,\right)\right)}\text{.}
\end{equation}

To eventually bound the min-entropy, only the first $\Omega$ elements from $\vec{\psi}$ and $\vec{\varsigma}$ are necessary.
$\Omega$ is the smallest $j$ such that $\sum_{i=0}^j |\varphi_i|\geq 2^{n_b}/|\randspace|$.
Due to \cref{eq:groupingCardinality}, $\Omega < J \ll 2^{n_b}/|\randspace|$.
Then the \emph{grouping bound} is
\begin{equation}
  H_\infty \left(X|Y\right) \leq -\log_2 \left(\vec{\psi} \exp_2\transp{\left(\vec{\varsigma}\,\right)}\right)\text{.}
\end{equation}

The resulting bounds for several $\theta_\Delta$ and codes, even for larger codes such as (127,8,31)-\textsc{bch}, are listed in \cref{tbl:entropies}.
Note that when setting $\theta_\tau$ to the lowest $p_i$ in a bias group $\rho_\tau$, the resulting estimate $H_\infty^{L} \left(X|Y\right)$ is no bound for entropy because the algorithm still uses the $2^{n_b}/|\randspace|$ most likely \gls{puf} responses.
However, for $H_\infty^H \left(X|Y\right)$ with $\theta_\tau$ set to the highest $p_i$ in a bias group, $H_\infty^{L} \left(X|Y\right)-H_\infty^H \left(X|Y\right)$ is an upper bound for the error made through the quantization of probabilities.

\emph{Remark:}
Until now we neglected to describe how to derive the sorted matrix $\check{Z}$, a task which is equivalent to computing how many bits have to be flipped from the bias groups $\rho_{\tau}$ for each of the $2^{n_b}/|\randspace|$ most likely \gls{puf} responses.
This is a non-trivial task and falls into the category of integer linear programming, which is NP-hard and would require to iterate over $J$ groups $\varphi_j$ in the worst case.

However, a function\footnote{\textsc{matlab} code available at \url{https://gitlab.lrz.de/tueisec/ind_puf_entropy}} can be written that, given a target range of probability, returns the flip vectors $\vec{\zeta_{j}}$ that result in such a probability by recursion within the $\rho_\tau$.
Once sorted, the returned vectors constitute a continuous part of $\check{Z}$.
The first $\Omega$ rows of $\check{Z}$ can thus be assembled from repeated calls to the function, starting with a target range from the probability of the best guess downwards, adding the results for consecutive target ranges until a sufficient part of $\check{Z}$ is obtained.

\section{Validating Entropy Bounds with Key Rank}\label{sec:keyrank}
The previous sections analyzed state-of-the-art methods and explained our extentions for the \gls{ind} case to estimate or lower bound the conditional min-entropy of a \gls{puf} response given the helper data.
We claimed that because the state-of-the-art tools only apply to the \gls{iid} case, users are forced to take this assumption, even if not justified, thereby jeopardizing the security of the system due to a severe overestimation of min-entropy.
The corresponding results are summarized in \cref{tbl:entropies}.

In this section, we aim to validate these results by simulating actual attacks on a code-offset construction using the $192$ devices that constitute the dataset from \cite{MaitiDataset}, which has been used throughout this work.
We randomly chose $10^4$ \SI{144}{\bit} keys (the maximum storable key length with the considered codes on our device) and implemented them on each device, using all codes mentioned in \cref{tbl:entropies}, to obtain corresponding helper data.
Because a device provides $256$ response bits, only $k=k_b\lfloor\frac{256}{n_b}\rfloor$ bits of the key can be stored for an $(n_b,k_b,t)$ \gls{ecc}.
Where $k<144$, we used the first $k$ bits of the key and the first $n=n_b\frac{k}{k_b}$ bits of the \gls{puf} response.
The entropy contained in these first $n$ response bits (not given any helper data) under the \gls{iid} and \gls{ind} assumption is reported in \cref{tbl:entropies} in columns $m$, $\tilde{m}$.
Given the helper data and the Bit-Alias $\vec{p}$, we determined the \emph{key rank} for each key, device, and \gls{ecc}.

The metric key rank originates from the field of side-channel analysis.
It represents the number of unsuccessful guesses an attacker would make until the correct key is found while following an optimal guessing strategy.
To find the optimal guessing strategy, one needs information on which key hypotheses are more likely. 
In side-channel analysis, such information results from e.g. measuring the electro-magnetic emanations of a device while it performs a cryptographic operation.
For \glspl{puf}, the Bit-Alias provides such information. 
An attacker may obtain a sufficiently accurate estimation of it by purchasing (as legitimate customer) a sufficient number of devices from the same type and analyzing them in detail.
To calculate the key rank even for \glspl{ecc} where $2^k$ is infeasible to enumerate, we use the method by Glowacz et al. \cite{Glowacz2015}.

\begin{figure}
  \centering
  \input{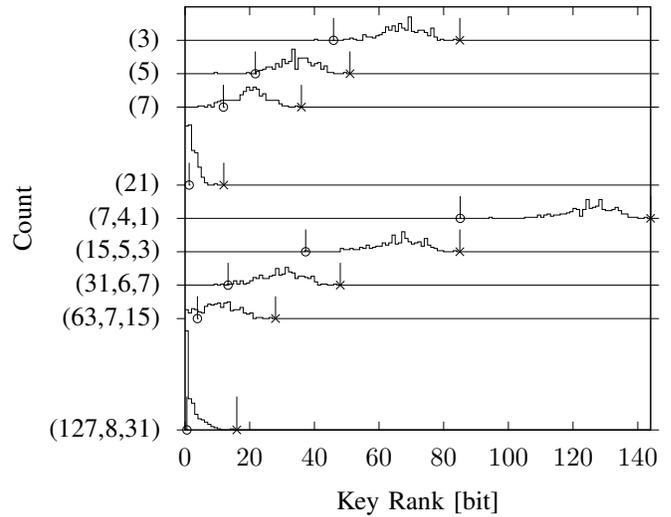}
  \caption{Histogram of key rank in \bit\ for all devices in the Maiti et al. dataset.
  Vertical lines indicate the grouping bound for $\theta_\Delta=0.05$ and the theoretical maximum $k$, e.g. $2 \cdot \SI{8}{\bit}=\SI{16}{\bit}$ for the (127,8,31) \textsc{bch}-code.}\label{fig:keyrank}
\end{figure}

\Cref{fig:keyrank} reports the logarithmic key rank in \bit\ for each device as a histogram for each \gls{ecc}.
The key rank depends on the device, but not the key, because the guessing strategy aims at the \gls{puf} response and once it is correctly guessed, the helper data provide a direct mapping to the key. 
Vertical bars with a cross indicate $k$, as $2^k$ is the number of all possible guesses for a \SI{k}{\bit} key.
Vertical bars with a circle indicate the result of the grouping bound for $\theta_\Delta=0.05$, i.e. a lower bound for $H_\infty (X|Y)$. 
According to \gls{nist}, $H_\infty - \SI{1}{\bit}$ provides a lower bound for average key rank if at most $k^\ast < k$ guesses are made per device \cite{NISTsp800-90B}.
Hence an attacker must expect to require at least $2^{H_\infty-1}$ guesses on average to find the correct key.

The validity of the bounds under \gls{iid} assumption is already falsified in \cref{tbl:entropies} by comparison to the exact average conditional min-entropy with our extension for \gls{ind}.
Comparison with the key rank shows that the (n-k) bound under \gls{ind} assumption is unpractically conservative.
The results for our grouping bound, however, are all safely below the average key rank without being too conservative.
This confirms that our work provides a valid bound to assess the security of a \gls{puf} based key storage using the common code-offset construction.

For the grouping bound itself, a trade off between accuracy and runtime complexity is observable:
Compared to the actual average conditional min-entropy, where it is feasible to compute, $\theta_\Delta=0.05$ provides the tighter bound.
However, it also requires a longer runtime due to the determination of the most likely flip vectors $\vec{\zeta_{j}}$:
For the computationally most expensive case of a (127,8,31) \textsc{bch}-code, the runtime on a commodity computer is $\approx \SI{20}{\minute}$ for $\theta_\Delta=0.05$, compared to $\approx \SI{10}{\second}$ for $\theta_\Delta=0.1$.
In either case, our grouping bound is the tightest bound under \gls{ind} assumption that is feasible for large codes.

\section{Conclusion}\label{sec:conclusion}
To verify the security of a \gls{puf} based key storage, the remaining conditional entropy of the key is crucial.
However, current methods are either inaccurate or infeasible without an \gls{iid} assumption for the \gls{puf} response.
Because the \gls{iid} assumption is not justified for many types of \gls{puf}, this work proposes an accurate and feasible method to lower bound the remaining entropy under less stringent \gls{ind} assumption.
Results of applying the bound to different codes and for real world data show the quality of the bound.
The results also demonstrate the relation between the bound and the average key rank, i.e. the effort an attacker must expect for guessing the key under an optimal guessing strategy.


\section*{Acknowledgment}
{\small This work was partly funded by the German Federal Ministry of Education and Research in the project \textsc{hqs} through grant number 16KIS0616.
Permanent \textsc{id} and revision date of this document:\\
3b0bccc028d2b79fcd67c668263243ce13b6112d

\isodate\printdate{2019-10-02}}



\bibliographystyle{IEEEtran}
\bibliography{IEEEabrv,abbreviateAuthors,docear,paper}
%
%
%

\end{document}